\documentclass[12pt,epsf]{article}
\usepackage{ulem}
\usepackage{color}
\usepackage{graphicx}
\usepackage{amsmath}
\usepackage{amsfonts}
\usepackage{cite}
\begin{document}

\def\a{\alpha}
\def\b{\beta}
\def\c{\varepsilon}
\def\d{\delta}
\def\e{\epsilon}
\def\f{\phi}
\def\g{\gamma}
\def\h{\theta}
\def\k{\kappa}
\def\l{\lambda}
\def\m{\mu}
\def\n{\nu}
\def\p{\psi}
\def\q{\partial}
\def\r{\rho}
\def\s{\sigma}
\def\t{\tau}
\def\u{\upsilon}
\def\v{\varphi}
\def\w{\omega}
\def\x{\xi}
\def\y{\eta}
\def\z{\zeta}
\def\D{\Delta}
\def\G{\Gamma}
\def\H{\Theta}
\def\L{\Lambda}
\def\F{\Phi}
\def\P{\Psi}
\def\S{\Sigma}
\def\hyphen{\mathchar`-}

\def\o{\over}
\def\beq{\begin{eqnarray}}
\def\eeq{\end{eqnarray}}
\newcommand{\gsim}{ \mathop{}_{\textstyle \sim}^{\textstyle >} }
\newcommand{\lsim}{ \mathop{}_{\textstyle \sim}^{\textstyle <} }
\newcommand{\vev}[1]{ \left\langle {#1} \right\rangle }
\newcommand{\bra}[1]{ \langle {#1} | }
\newcommand{\ket}[1]{ | {#1} \rangle }
\newcommand{\EV}{ {\rm eV} }
\newcommand{\KEV}{ {\rm keV} }
\newcommand{\MEV}{ {\rm MeV} }
\newcommand{\GEV}{ {\rm GeV} }
\newcommand{\TEV}{ {\rm TeV} }
\def\diag{\mathop{\rm diag}\nolimits}
\def\Spin{\mathop{\rm Spin}}
\def\SO{\mathop{\rm SO}}
\def\O{\mathop{\rm O}}
\def\SU{\mathop{\rm SU}}
\def\U{\mathop{\rm U}}
\def\Sp{\mathop{\rm Sp}}
\def\SL{\mathop{\rm SL}}
\def\tr{\mathop{\rm tr}}
\def\mpl{M_{PL}}

\def\IJMP{Int.~J.~Mod.~Phys. }
\def\MPL{Mod.~Phys.~Lett. }
\def\NP{Nucl.~Phys. }
\def\PL{Phys.~Lett. }
\def\PR{Phys.~Rev. }
\def\PRL{Phys.~Rev.~Lett. }
\def\PTP{Prog.~Theor.~Phys. }
\def\ZP{Z.~Phys. }


\baselineskip 0.7cm

\begin{titlepage}

\begin{center}
\hfill IPMU-15-0072 \\
\hfill \today

\vskip 3cm

{\large\bf Mass-Splitting between Haves and Have-Nots
}\\
-- Symmetry vs. Grand Unified Theory --

\vskip 1.2cm
Keisuke Harigaya$^1$,
Masahiro Ibe$^{1,2}$
and
Motoo Suzuki$^{1,2}$
\vskip 0.4cm
$^1${\it ICRR, University of Tokyo, Kashiwa 277-8582, Japan}\\
$^2${\it Kavli IPMU (WPI), UTIAS, University of Tokyo, Kashiwa 277-8583, Japan}
\vskip 1.5cm

\abstract{
We revisit the long-standing problem of supersymmetric grand unified theory (GUT), 
the doublet-triplet splitting problem.
We discuss
whether symmetry
which controls the $\mu$
term
in the minimal supersymmetric standard model
is compatible with GUT.
We find that the symmetry
must be broken at the GUT scale.
A similar argument also shows that
the $R$ symmetry, which is important for low energy supersymmetry, must be broken down to a $Z_{2R}$ symmetry at the GUT scale.
We propose a new prescription to achieve the doublet-triplet splitting by symmetry.
There, the symmetry
which controls the $\mu$ term
is spontaneously broken at the GUT scale by order parameters which are 
charged under other symmetries.
Bilinear terms of triplet Higgses are {\it charged} under the other symmetries, 
while those of doublet Higgses are {\it not}.
Then triplet Higgses directly couple to the order parameters and hence obtain GUT scale masses,
while doublet Higgses obtain suppressed masses.
The broken $R$ symmetry can be also effectively preserved by a similar prescription.
As a demonstration,
we construct an $SU(5)\times SU(5)$ GUT model.
We also comment on unification of yukawa couplings.
}
\end{center}
\end{titlepage}

\section{Introduction}
For decades, supersymmetry (SUSY) has been expected to be an important key 
for physics beyond the Standard Model (SM). 
In particular, the supersymmetric standard model (SSM) has been thought to be
very successful since  it allows a vast separation of low energy scales from 
high energy scales~\cite{Maiani:1979,Veltman:1980mj,Witten:1981kv,Kaul:1981wp}
such as the Planck scale or the scale of the Grand Unified Theory (GUT)~\cite{Georgi:1974sy}.
The SSM has been also supported
by unification of the gauge coupling constants 
of the SM at around $10^{16}$\,GeV.
Interesting connection between the proton stability and the stability of a dark matter candidate, 
the lightest supersymmetric particle (LSP),
also 
illuminates the success of the SSM.

With
the Large Hadron Collider (LHC) showing no evidence for SUSY~\cite{lhc_susy}, 
however,
the situation for natural electroweak symmetry breaking
by SUSY particles around a sub TeV scale
has grown increasingly severe.
Besides, the observed Higgs boson mass of 125\,GeV~\cite{lhc_higgs} 
seems to point the masses of SUSY particles in a tens to hundreds TeV range in the minimal SSM (MSSM).
These facts might suggest that it is more difficult to obtain immediate hints on the SSM from collider
experiments than anticipated before the LHC experiments.

With this little gloomy outlook, it is worthy to reappraise one of the strongest motivations of the SSM,
the successful GUT, and try to obtain implications on the structure of the SSM.
In particular, 
it is important to think again 
a strong correlation between the $\mu$-problem in the MSSM
and the doublet-triplet splitting problem in the SUSY GUT~\cite{Dimopoulos:1981zb,Witten:1981kv,Sakai:1981gr}.
On the one hand, a variety of solutions to the doublet-triplet splitting problem have been proposed so far, such as  
missing partner mechanism~\cite{Masiero:1982fe,Grinstein:1982um},
missing vacuum expectation value (VEV) mechanism~\cite{missvev},
product GUT models~\cite{Yanagida:1994vq},
and orbifold GUT models~\cite{Kawamura:2000ev,Hall:2001pg}. 
On the other hand, many successful models of the MSSM have been 
proposed in which the size of the $\mu$-term is controlled by a symmetry such 
as the $R$ symmetry~\cite{Inoue:1991rk,Casas:1992mk} or the Peccei-Quinn symmetry~\cite{Kim:1983dt}.%
\footnote{The symmetry which forbids the $\mu$-term is also important to 
suppress the so-called the dimension five proton decay operators.} 
Although these two problems are intimately related with each other, 
the solutions to these problems are often discussed separately.

In this paper, we discuss these issues with a special emphasis on the 
consistency between a symmetry which controls the $\mu$-term and solutions to the 
doublet-triplet splitting problem.
In fact, it has been shown in Refs.\,\cite{Goodman:1985bw,Witten:2001bf} that the low energy symmetry which forbids the $\mu$-term at the GUT scale cannot be consistently embedded in GUT models when 
the SM gauge groups are embedded into a simple GUT group (see also Ref.~\cite{Fallbacher:2011xg}).
As we will see, this no-go theorem can be extended to GUT models based on a
product group such as $SU(5)\times SU(5)$ where the SM gauge groups are embedded into the GUT group
so that the coupling unification is automatically maintained.%
\footnote{If the SM gauge groups are embedded into an asymmetric product gauge groups 
such as $SU(5)\times U(3)$, it is actually possible to embed the low energy symmetry which forbids 
the $\mu$-term consistently to the GUT~\cite{Yanagida:1994vq,Izawa:1997he}. 
In such models, however, the coupling unification is not automatically achieved and 
requires that the gauge couplings other than that of $SU(5)$ are strong at the GUT scale.
}
It should be noted that
this extension is non-trivial since the low energy symmetry does not necessarily 
commute with the GUT group in the case of a product  GUT.

After discussing the no-go theorem, we discuss how to mend the low energy 
symmetry in the MSSM which forbids the $\mu$-term 
with GUT models with automatic coupling unification.
There, we show a prescription such that the symmetry which forbids 
the mass of the doublet (hereafter, we refer to this symmetry as ``doublet symmetry")
is broken at the GUT scale while its breaking does not generate the $\mu$-term of the GUT scale.
Concretely, we assume that the order parameters of the doublet symmetry are also charged 
under other symmetries under which the doublets are not charged.
Then, if the color triplet Higgses are charged under the above other symmetries, 
the triplet can obtain the mass of the GUT scale (hereafter, we refer to the other symmetries as
the ``triplet symmetries") while the doublet Higgs mass is suppressed due to the lack 
of the charges of the triplet symmetries.
It should be emphasized that this mechanism is close but opposite to the ``collective symmetry breaking",
where Lagrangian terms charged under multiple symmetries are more suppressed.
In our mechanism, instead, less charged fields obtain more suppressed masses, i.e.,
the haves get large masses while the have-nots get no masses at the GUT scale.

We also find that the $R$ symmetry, which is important for low energy SUSY,  
should be broken down to $Z_{2R}$ symmetry at the GUT scale 
when the coupling unification is guaranteed.
In general, such a large breaking leads to a large VEV of the superpotential,
which is incompatible with low energy SUSY.
This worry can be easily solved if the order parameters of the $R$ symmetry
are also charged under other symmetries, as in the case of the doublet symmetries. 

By observing the similar requirements  on the doublet symmetry and the $R$ symmetry,  
we demonstrate GUT model building where the doublet symmetry is the $R$ symmetry.
As 
order parameters of the $R$ symmetry are charged under triplet symmetries,
the $\mu$-term is not generated at the GUT scale.	
The breaking of the $R$ symmetry generates a constant term in the superpotential
which is suppressed by triplet symmetries.
Eventually, the VEV of the superpotential leads to the $\mu$-term of the order of the gravitino mass, $m_{3/2}$,
via Planck-suppressed operators~\cite{Inoue:1991rk,Casas:1992mk}.
As a bonus of our concrete example, we find an interesting connection between the gravitino mass 
and the GUT scale.
We also show that the infamous problem of the unification of down-quark yukawa couplings and charged-lepton yukawa couplings can be solved rather simply in our example.

This paper is organized as follows.
In Sec.~\ref{sec:splitting}, we prove the above mentioned no-go theorem on a
symmetry which forbids the $\mu$-term. 
We also propose a prescription to achieve the doublet-splitting by using triplet symmetries.
In Sec.~\ref{sec:RGUT}, we discuss the consistency between the $R$ symmetry and unification.
In Sec.~\ref{sec:model}, we construct a concrete realization of the mechanism.
In Sec.~\ref{sec:masses}, we discuss the detailed vacuum structure and the mass
spectrum of the model constructed in  Sec.~\ref{sec:model}.
The final section is devoted to conclusions and discussion.

\section{Mass Splitting and Unification}
\label{sec:splitting}
\subsection{Doublet symmetry and coupling unification}
In this subsection, we discuss the consistency between a low energy symmetry which forbids the $\mu$-term
in the MSSM
and the GUT gauge symmetry.
Let us refer to the low energy symmetry which is embedded in the GUT as the doublet symmetry.
Then, if the doublet symmetry remains unbroken at the GUT scale, the doublet mass is generated 
only after the remaining doublet symmetry is broken at a scale well below the GUT scale. 
If the triplet Higgs mass is, on the other hand, allowed by the doublet symmetry in some way, they
obtain the mass of the GUT scale.
This is the situation assumed in successful models of the MSSM where the $\mu$-problem
is solved by symmetries.
Unfortunately, however, we will immediately see that this possibility is incompatible 
with GUT models where the coupling unification is automatically maintained.

Since we are interested in GUT models which exhibit automatic coupling unification,
let us first discuss a $SU(5)$ GUT model as an example.
The following arguments can be extended to models with more generic simple GUT gauge groups.
Throughout  this paper, we assume that the doublet Higgses are placed in chiral supermultiplets  $H$ and $\bar{H}$
transforming ${\mathbf 5}\oplus{\mathbf{\bar 5}}$.
This choice is quite natural since it allows the yukawa interactions in the 
MSSM easily embedded in the GUT model where quarks and leptons are 
unified into chiral supermultiplets transforming ${\mathbf{\bar 5}}\oplus{\mathbf{10}}$.

Now, let us show that the doublet-triplet splitting with the unbroken doublet symmetry is impossible
under the reasonable assumption:
the Higgs multiplets are not mixed with quark and lepton multiplets in the GUT representations.
As we have mentioned earlier,
we have discarded the possible mixing
since such a complicated structure makes it difficult to obtain 
appropriate yukawa interactions while keeping the proton stability etc.
Under this assumption, we can discuss the anomaly 
matching condition of the doublet symmetry only within the Higgs sector.

We normalized the sum of the charges of a pair of doublet Higgses as $2q$.%
\footnote{When the doublet symmetry is the $R$ symmetry, charges we discuss 
denote those of fermion components.}
Below the GUT scale, only the doublet Higgses contribute to the anomaly of the doublet symmetry, $D$.
Thus, the contributions of the Higgs sector to $D\hyphen SU(3)_c\hyphen SU(3)_c$ and  $D\hyphen SU(2)_L\hyphen SU(2)_L$
are given by
\begin{eqnarray}
\label{eq:anom0}
A_{D\hyphen SU(3)_c\hyphen SU(3)_c}^{\rm Higgs} = 0,~~
A_{D\hyphen SU(2)_L\hyphen SU(2)_L}^{\rm Higgs} = 2q,
\end{eqnarray}
and hence, they are not equal with each other.
If the doublet symmetry is unbroken, on the other hand, the anomaly matching condition leads to
\begin{eqnarray}
A_{D\hyphen SU(3)_c\hyphen SU(3)_c}^{\rm Higgs} = A_{D\hyphen SU(2)_L\hyphen SU(2)_L}^{\rm Higgs} 
= A_{D\hyphen SU(5)\hyphen SU(5)}^{\rm Higgs} \ ,
\end{eqnarray}
which is not consistent with Eq.\,(\ref{eq:anom0}).
Therefore, we find that the unbroken doublet symmetry is incompatible with the doublet-triplet splitting in 
the $SU(5)$ GUT model.
This arguments can be easily extended to GUT models based on simple GUT groups 
such as $SO(10)$ or larger groups which contain a unique $SU(5)$ subgroup to which the SM gauge 
groups are embedded in.
In such cases, the above arguments can be repeated by using the $SU(5)$ subgroup. 

Next, let us consider models with product GUT gauge groups.
Even in such cases, the coupling unification is also automatically maintained
when the SM gauge groups originate from a {\it single simple} group.
For example, in $SU(5)_1 \times SU(5)_2$ models, $SU(3)_c$, $SU(2)_L$ and $U(1)_Y$ originate from a vector 
$SU(5)$ part of $SU(5)_1 \times SU(5)_2$, which leads to automatic coupling unification.
Unfortunately, however, things are no different in this class of product GUT models as for the no-go theorem.
That is, as we have shown, $A_{D\hyphen SU(3)_c\hyphen SU(3)_c}^{\rm Higgs}\neq A_{D\hyphen SU(2)_L\hyphen SU(2)_L}^{\rm Higgs}$ if the doublet-triplet splitting occurs successfully.
Then, the anomaly matching condition of the doublet symmetry and the GUT gauge group (in particular the vector 
$SU(5)$ part) again contradicts with $A_{D\hyphen SU(3)_c\hyphen SU(3)_c}^{\rm Higgs}\neq A_{D\hyphen SU(2)_L\hyphen SU(2)_L}^{\rm Higgs}$.
It should be noted the anomaly matching should be satisfied even when
the low energy doublet symmetry
is a diagonal unbroken part of
a symmetry and
a subgroup of $SU(5)_1\times SU(5)_2$
which do not commute with the SM gauge groups.%
\footnote{The discussion in \cite{Goodman:1985bw,Witten:2001bf} cannot be used when the low energy doublet symmetry
contains
a subgroup of $SU(5)_1\times SU(5)_2$,
since their arguments are based on the Weyl symmetry of the characters of representations 
for a fixed charges of the doublet symmetry.
}
This is because the subgroup of $SU(5)_1\times SU(5)_2$ is anomaly-free.
Therefore, we again find that the unbroken doublet symmetry is incompatible with the doublet-triplet splitting
in a class of product  GUT models where the SM gauge groups are embedded in a single simple group.%
\footnote{In product GUT models such as $SU(5)\times U(3)_H$, the difference of the anomalies does not 
lead to any inconsistency, 
since $SU(3)_c$ is a linear combination of $SU(3)\supset SU(5)$ and $SU(3)_H$ while $SU(2)_L\supset SU(5)$.
Even in this case, if the product group is embedded in a simple group eventually,
the above no-go theorem holds.
}

Before closing our discussion, let us comment on possible ways to evade the no-go theorem.%
\footnote{
For simple GUT gauge groups, the discussion in Refs.~\cite{Goodman:1985bw,Witten:2001bf} excludes the first possibility. 
}
\begin{itemize}
\item {\bf Higgs-matter mixing:}
If Higgs multiplets mix with quarks and leptons multiplets, the above anomaly argument must involve quarks 
and leptons, and hence, we might be able to evade the no-go-theorem.
It would not be easy to obtain appropriate yukawa couplings of quarks and leptons.
Also, it is non-trivial whether the $R$-parity can be conserved.
\item {\bf Accidentally light additional field:}
If there are GUT incomplete light multiplets
(we refer them surplus multiplets)
with non-trivial charges under 
the doublet symmetry, the anomaly mismatching can be evaded.
However, those multiplets ruin the coupling unification, generically.
If there are additional GUT incomplete multiplets, whose masses are allowed by the doublet symmetry but
accidentally as large as the masses of the surplus multiplets, and they fit into GUT complete multiplets,
the coupling unification is maintained.
However, the lightness of the additional GUT incomplete multiplets brings up the mass splitting problem again.
\item {\bf Explicite Breaking of the doublet symmetry by strong dynamics:}
If the doublet symmetry is explicitly broken, the above anomaly argument is invalidated.
In $SU(5)\times SU(5)$ models presented in Ref.~\cite{Izawa:1997br,Kitano:2001ie}, the classical doublet symmetry has anomaly of a hidden strong gauge interaction.
A dynamically generated superpotential explicitly breaks the classical doublet symmetry.
The doublet mass is, however, absent due to missing VEVs.
The doublet mass is given by the breaking of the classical doublet symmetry at a low energy scale.
\end{itemize}

\subsection{Broken doublet symmetry and use of triplet symmetry}
\label{sec:mechanism}
In this paper, we take a different approach.
We break the doublet symmetry {\it at the GUT scale.}
Then the above mentioned no-go theorem, which is based on the low energy theory with
unbroken doublet symmetry, 
does not hold anymore.

The doublet symmetry broken at the GUT scale, however, seems to lead to a doublet mass around the GUT scale.
To avoid the generation of the doublet mass at around the GUT scale,
we assume that the order parameters of  the doublet symmetry
are
charged under some 
other symmetries.
We refer to those additional symmetries as ``triplet symmetries".
The triplet Higgses (and other unwanted fields) are charged under the triplet symmetries appropriately.
Then the order parameters of the doublet symmetry may directly couple to the triplet Higgses
and give GUT scale masses to them. 
On the other hand, if the doublet Higgses  are neutral under the triplet symmetries,
 direct couplings between the order parameters and the doublet Higgses  are forbidden.%
\footnote{
The triplet symmeties may controll flavor structure of quarks and leptons,
if the doublet Higgses  are charged under them,
while keeping the doublet-triplet splitting~\cite{Antusch:2014poa}.}
The doublet mass is given only by higher dimensional operators and hence is suppressed in comparison with the GUT scale:~the doublet symmetry is effectively preserved as a good approximate symmetry below the GUT scale, 
with an aid of the triplet symmetries.%
\footnote{The doublet-triplet splitting with broken doublet symmetry
has been also considered in Ref.\,\cite{Dine:2002se}.}

This prescription should be compared with the usual collective symmetry breaking mechanism 
in which Lagrangian terms charged under multiple symmetries are more suppressed than 
the less charged ones.
In our prescription, instead, the masses of less charged fields are more suppressed.
That is, the {\it haves} get large masses while the {\it have-nots} get no masses at the GUT scale.
From the view point of the low energy MSSM models, this mechanism is advantageous,
since the $\mu$-term looks controlled only by the doublet symmetry, and hence,
many successful ideas to control the $\mu$-term by a symmetry can be embedded in
GUT models without changing the symmetry structure of the low energy MSSM.

\section{$R$ symmetry and Unification}
\label{sec:RGUT}
In low energy SUSY, the VEV of the superpotential, $W_0$, is
required to be small in order to achieve an almost vanishing cosmological constant.
Such a small VEV can be achieved if there is a symmetry which prevents 
$W_0$ from being very large, namely an $R$ symmetry. 
In this section, let us comment on the consistency between the $R$ symmetry and unification.
We immediately see that the $R$ symmetry should be broken at the GUT 
scale in GUT models with automatic coupling unification.

Now, let us repeat the above arguments of the anomaly matching 
in the case of the $R$ symmetry.
In the MSSM gaugino sector, the anomalies of the $R$ symmetry are given by,
\begin{eqnarray}
\label{eq:anomR}
A_{R\hyphen SU(3)_c\hyphen SU(3)_c}^{\rm Gaugino} = 6,~~
A_{R\hyphen SU(2)_L\hyphen SU(2)_L}^{\rm Gaugino} = 4.
\end{eqnarray}
If the $R$ symmetry is unbroken, on the other hand, the anomaly matching condition leads to 
\begin{eqnarray}
\label{eq:anomRGUT}
A_{R\hyphen SU(3)_c\hyphen SU(3)_c}^{\rm Gaugino} = A_{R\hyphen SU(2)_L\hyphen SU(2)_L}^{\rm Gaugino} 
= A_{R\hyphen SU(5)\hyphen SU(5)}^{\rm Gaugino + GUT\mbox{-}breaking} \ .
\end{eqnarray}
Here, $SU(5)$ denotes a simple subgroup of the GUT gauge group 
in which the standard model gauge groups are embedded in.
It should be noted that the GUT breaking sector contributes to the matching condition since 
the mass partners of heavy GUT gauge multiplets are the would-be Nambu-Goldstone modes.
It should be also noted that the multiplets in the GUT breaking sector 
never mix the quark, the lepton nor the Higgs sectors.
Thus, the quark, the lepton nor the Higgs sectors do not contribute to the above condition.
We find that the anomalies in Eq.\,(\ref{eq:anomR}) are consistent with the condition in Eq.\,(\ref{eq:anomRGUT})
only when the $R$ symmetry is broken down to $Z_{2R}$ at the GUT scale.

This no-go theorem shows that it is not easy to embed a low energy theory with an $R$ symmetry 
into GUT models. 
As  in the case of the doublet symmetry, one way to make the low energy $R$ symmetry compatible 
with GUT models is to assume that the order parameters of the $R$ symmetry are charged under other symmetries.
In this case, if the low energy fields are  neutral under the other symmetries, the $R$ symmetry 
is effectively preserved as a good approximate symmetry below the GUT scale.

The other symmetries are also helpful to suppress the VEV of the superpotential,
so that the gravitino mass is small enough to be compatible with low energy SUSY.
In the concrete model in the following section, we obtain the gravitino mass  
as small as $O(10^2-10^6)$ GeV (see Fig.~\ref{fig:m32}), which is consistent with gravity mediation.
It seems to be, however, not easy to obtain a gravitino mass appropriate for low energy gauge mediation.

\section{$R$ symmetry as  doublet symmetry}
\label{sec:model}
As we have discussed above, the doublet symmetry as well as the $R$ symmetry 
are required to be broken at the GUT scale.
If the order parameters of these symmetries are charged under other symmetries, 
they are effectively preserved below the GUT scale.
By observing this similarity, it is quite enticing to identify the $R$ symmetry as a doublet symmetry. 

In fact, the use of $R$ symmetry to control the $\mu$-term in the MSSM is considered to be 
one of the most successful solution to the $\mu$-problem~\cite{Inoue:1991rk,Casas:1992mk}.
When the pair of doublet Higgses  has a vanishing $R$-charge (i.e.~charge $-2$ in terms of the charge of fermionic components),
the $\mu$-term of the gravitino mass size is naturally generated when the doublet Higgses
couples to the VEV of the superpotential.
It should be  emphasized that  this solution works without having singlet
SUSY breaking fields unlike the usual Giudice-Masiero mechanism~\cite{Giudice:1988yz}.
This mechanism is, therefore, particularly suitable for a class of high scale SUSY 
breaking models~\cite{Ibe:2006de,Ibe:2011aa,Ibe:2012hu,Bhattacherjee:2012ed}
with a gravitino mass in hundreds to thousands TeV range.
Gaugino masses are dominantly given by the anomaly mediation~\cite{AMSB} (see also 
\cite{Bagger:1999rd, D'Eramo:2013mya, Harigaya:2014sfa}),
where no singlet SUSY breaking fields are required.%
\footnote{The absence of singlet SUSY breaking fields is advantageous by itself in view of the 
so-called Polonyi problem~\cite{Coughlan:1983ci,Ibe:2006am,Harigaya:2013ns}.}

\subsection{$R$ symmetry breaking and  gravitino mass}
We consider a discrete $R$ symmetry, $Z_{2nR} (n>1)$,   as the doublet symmetry.
We assume that the discrete $R$ symmetry $Z_{2nR}$ is spontaneously broken by the following superpotential,
\begin{eqnarray}
W = v^n \phi - \frac{\lambda}{n+1}\phi^{n+1},
\end{eqnarray}
where $v^n$ and $\lambda$ are constants and $\phi$ is a chiral multiplet.
Here and hereafter, we take the reduced Planck mass to be unity.

The charge assignments of $\phi$ and $v^n$ are as given in Table~\ref{tab:charge1}.
There, we introduce a discrete symmetry $Z_{n+1}$ along with $Z_{2nR}$ symmetry, 
which will be identified with a part of triplet symmetries in the later discussion.
The constant, $v^n$, is a spurious field of the breaking of $Z_{n+1}$ which may be considered to be generated 
dynamically. 

\begin{table}[t]
\caption{\sl \small Charge assignment of chiral multiplets. 
The charges of $\phi$ and $v^n$ are uniquely determined by the 
definitions of the symmetries, $Z_{2nR}$ and $Z_{n+1}$.
The vanishing charges of $H_{1}$ and $\bar H_{1}$ are the simplest 
implementations of the $Z_{2nR}$ as the doublet symmetry and $Z_{n+1}$ as the triplet symmetry.
The charge assignments of $\Phi_{2}$ and $\bar\Phi_2$ are generic at this point.
Since we eventually allow the Higgs bi-linear terms in Eq.\,(\ref{eq:DTW}),
the charges of $H_2$ and $\bar{H}_2$ and $\bar{\Phi}_3$ 
are subsequently determined for given charges of $\Phi_3$
and a given value of $m =1,2$.}
\begin{center}
\begin{tabular}{|c|cccccccccc|}
\hline
& $\phi$ & $v^n$ &  $\Phi_3$ & $\Phi_2$ & $\bar{\Phi}_3$ & $\bar{\Phi}_2$ & $H_1$ & $\bar{H}_1$ & $H_2$ & $\bar{H}_2$ \\
\hline
$Z_{2nR}$ & 2 & 0         & $r_3$ & $r_2$  & $2+2m-r_3$  & $r_{\bar{2}}$  &0&0&$2-r_3$& $r_3 - 2m$ \\
$Z_{n+1}$ & $-1$ & 1   & $q_3$ & $q_2$  & $-m-q_3$  & $q_{\bar{2}}$  &0&0&$-q_3$& $q_3 + m$\\
\hline
$SU(5)_1$ &${\bf 1}$&${\bf 1}$& ${\bf 5}$ & ${\bf 5}$ & ${\bf \bar{5}}$ & ${\bf \bar{5}}$ &${\bf 5}$& ${\bf \bar{5}}$&${\bf 1}$&${\bf 1}$\\
$SU(5)_2$ &${\bf 1}$&${\bf 1}$&  ${\bf \bar{5}}$ & ${\bf \bar{5}}$ & ${\bf 5}$ & ${\bf 5}$&${\bf 1}$&${\bf 1}$&${\bf 5}$&${\bf \bar{5}}$\\
\hline
\end{tabular}
\end{center}
\label{tab:charge1}
\end{table}

\begin{figure}[t]
\begin{center}
  \includegraphics[width=0.6\linewidth]{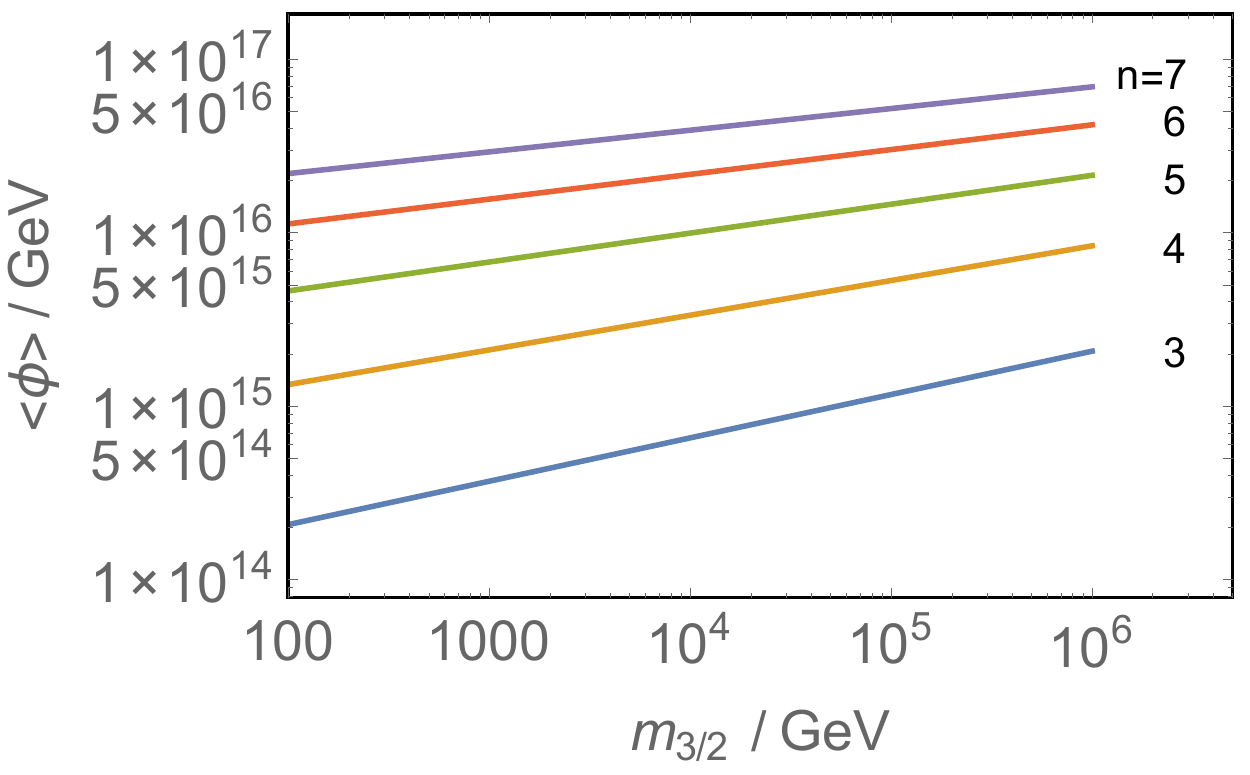}
 \end{center}
\caption{\sl \small
The VEV of $\phi$ for a given gravitino mass $m_{3/2}$ for a given order of the discrete $R$ symmetry, $Z_{2nR}$.
Here, we have taken $\lambda=1$.
}
\label{fig:m32}
\end{figure}

At the vacuum, the $R$ symmetry is spontaneously broken down to $Z_{2R}$ symmetry.
The VEV of $\phi$ and that of the superpotential are given by
\begin{eqnarray}
\vev{\phi} = \lambda^{-1/n} v,~~~
W_0\equiv \vev{W} =  \frac{n}{n+1}\lambda^{-1/n}v^{n+1}
= \frac{n}{n+1}\lambda \vev{\phi}^{n+1}.
\end{eqnarray}
In Fig.~\ref{fig:m32}, we show the relation between the gravitino mass $m_{3/2} = W_0$ and $\vev{\phi}$
for $n=3\mathchar`-7$. Here, we take $\lambda=1$.
It should be noted that $\vev{\phi}$ is around the GUT scale $\sim 10^{16}$ GeV
for a wide range of $m_{3/2}$ with $n=5$-$7$.
Since the $R$ symmetry is broken down to $Z_{2R}$, the $R$ symmetry itself does not
forbid the doublet Higgs mass of the GUT scale anymore.
As we demonstrate below, the triplet symmetries instead forbid the doublet Higgs mass although the doublet Higgses
are {\it neutral} under the triplet symmetries. 
Eventually, the MSSM have a discrete $R$ symmetry as a 
good approximate symmetry.

\subsection{$SU(5)_1\times SU(5)_2$ breaking}
To implement the triplet symmetries, let us consider an $SU(5)_1\times SU(5)_2$ GUT gauge group~\cite{Barbieri:1994jq} 
as an example,
where, as we will see shortly, the doublet-triplet splitting is achieved rather easily.%
\footnote{The use of the product is not mandatory to implement the prescription in Sec~\ref{sec:mechanism}.
}
To break  $SU(5)_1\times SU(5)_2$ down to $SU(3)_c\times SU(2)_L \times U(1)_Y$,
we introduce four chiral multiplets in the bi-fundamental representation, 
 $\Phi_3$, $\Phi_2$, $\bar{\Phi}_3$ and $\bar{\Phi}_2$ (see Table~\ref{tab:charge1})
which transform under $SU(5)_1\times SU(5)_2$ as,
\begin{eqnarray}
\Phi \rightarrow e^{i \alpha_1^I T^I}\Phi e^{-i \alpha_2^I T^I},~~
\bar{\Phi} \rightarrow e^{i \alpha_2^I T^I}\bar \Phi e^{-i \alpha_1^I T^I}.
\end{eqnarray}
Here, $T^I (I = 1\hyphen24)$ are generators of $SU(5)$
and $\alpha_1^I$ and $\alpha_2^I$ are parameters of $SU(5)_1$ and $SU(5)_2$ transformations, respectively.
Specifically,
\begin{eqnarray}
T^{1\hyphen8} = 
\begin{pmatrix}
\lambda^{a=1\hyphen8} & 0\\
0 & 0
\end{pmatrix},~~
T^{9\hyphen11} = 
\begin{pmatrix}
0 &0  \\
0 & \frac{1}{2}\sigma^{i=1\hyphen 3}
\end{pmatrix},~~
T^{24}=
\frac{\sqrt{15}}{30}
\begin{pmatrix}
2\times {\bf 1}_3 & 0\\
0 & -3 \times {\bf 1}_2
\end{pmatrix},
\end{eqnarray}
where $\lambda^a$ are Gell-Mann matrices and  $\sigma^i$ are  Pauli matrices.
We assume that the vacuum with $SU(3)_c\times SU(2)_L \times U(1)_Y$ symmetry is achived
by the following VEVs,
\begin{eqnarray}
\label{eq:VEV}
\vev{\Phi_3} =  v_3 
\begin{pmatrix}
{\bf 1}_3 & 0\\
0 & 0
\end{pmatrix},~~
 \vev{\bar{\Phi}_3} = \bar{v}_3 
\begin{pmatrix}
{\bf 1}_3 & 0\\
0 & 0
\end{pmatrix},~~
\vev{\Phi_2} =  v_2
\begin{pmatrix}
0 &0  \\
0 & {\bf 1}_2
\end{pmatrix},~~
 \vev{\bar{\Phi}_2} = \bar{v}_2
\begin{pmatrix}
 0&0  \\
0 & {\bf 1}_2
\end{pmatrix}, 
\end{eqnarray}
where ${\bf 1}_{\ell}$ denotes the  $\ell$-dimensional unit matrix.
The remaining $SU(3)_c$, $SU(2)_L$ and $U(1)_Y$ transformations are given by parameters
$\alpha_1^{1\hyphen8}= \alpha_2^{1\hyphen 8}$, $\alpha_1^{9\hyphen11}= \alpha_2^{9\hyphen 11}$ and $\alpha_1^{24}= \alpha_2^{24}$, respectively.
As long as $v_3\sim v_2$, the coupling constants of $SU(3)_c$, $SU(2)_L$ and $U(1)_Y$ are unified around the scale $v_2\sim v_3$, since the SM gauge groups are embedded in a single vector $SU(5)$ subgroup of $SU(5)_1\times SU(5)_2$.
The stabilization of the VEVs is discussed in the next section.
There, we obtain $v_2\sim v_3 \sim \vev{\phi}$ in a natural way.

Let us  mention a symmetry peculiar to $SU(5)_1\times SU(5)_2$ models, which is quite useful 
to achieve the doublet-triplet splitting.
Consider transformations with $\alpha_1^{24} = - \alpha_2^{24}$, which we refer to 
as $U(1)_{24}^A$ transformations.
Under $U(1)_{24}^A$ symmetry,
$v_3$, $\bar{v}_3$, $v_2$ and $\bar{v}_2$
have charges of
$4$, $-6$, $-4$ and $6$, respectively.%
\footnote{Here, the normalization of the rotation angle of $U(1)_{24}^A$ is 
$\sqrt{15}/30$ times different from those of $\alpha^{24}$'s.
}
A triplet and a doublet in ${\bf 5}$ of $SU(5)_1$ have $U(1)_{24}^A$ charges of  $2$ and $-3$, while
those of $SU(5)_2$ have $U(1)_{24}^A$ charges of  $-2$ and $+3$.

\subsection{Doublet-triplet splitting}
In our discussion, we assume that quarks and leptons are unified into ${\bf \bar 5}$ and ${\bf {10}}$ of $SU(5)_1$.
We also assume that the doublet Higgses  $H_u$ and $H_d$ are placed in ${\bf 5}$ and ${\bf \bar{5}}$ of $SU(5)_1$,
which we denote by $H_1$ and $\bar{H}_1$;
\begin{eqnarray}
H_1 =\begin{pmatrix}
H_1^T\\
H_u
\end{pmatrix},~~
\bar{H}_1 =\begin{pmatrix}
\bar{H}_1^T\\
H_d
\end{pmatrix} .
\end{eqnarray}
With this arrangement, we can easily obtain yukawa couplings of quarks and leptons without any suppression,
such as $H_1\, {\bf 10}\, {\bf 10}$ and $H_2\, {\bf \bar{5}} \,{\bf 10}$.

Note that combinations $H_1^T \bar{H}_1^T$ and $H_u H_d$ have identical charges under any symmetries.
Thus, it is impossible to achieve hierarchy between their masses without fine-tuning
if we assume that the triplet mass comes from the bi-linear term of $H_1^T$ and $\bar{H}_1^T$.
To avoid this problem, we are lead to introduce a pair of ${\bf 5}$ and ${\bf \bar{5}}$ of $SU(5)_2$, 
which we denote as $H_2$ and $\bar{H}_2$;
\begin{eqnarray}
H_2 =\begin{pmatrix}
H_2^T\\
H_2^D
\end{pmatrix},~~
\bar{H}_2 =\begin{pmatrix}
\bar{H}_2^T\\
\bar{H}_2^D
\end{pmatrix}.
\end{eqnarray}
The $U(1)_{24}^A$ charges of these Higgses are given by
\begin{eqnarray}
H_1^T: 2,~
\bar{H}_1^T:-2,~
H_2^T: -2,~
\bar{H}_2^T:2,\nonumber \\
H_u : -3,~
H_d: 3,~
H_2^D: 3,~
\bar{H}_2^D:-3.
\end{eqnarray}
The charge assignment of those Higgsses is given in Table~\ref{tab:charge1}.
It should be noted that the mass terms of the Higgs multiplets are forbidden by $Z_{2nR}$ symmetry.
In particular, the bi-linear term $H_1\bar H_1$ is forbidden {\it only} by the $Z_{2nR}$ symmetry,
and hence, it is identified with the doublet symmetry.

\begin{table}[t]
\caption{\small \sl $U(1)_{24}^A$, $Z_{2nR}$ and $Z_{n+1}$ charges of Higgs quadratic terms.
These charge assignments can be read off from Table.\,\ref{tab:charge1}.
}
\begin{center}
\begin{tabular}{|c|cccc|}
\hline
 & $H_1^T \bar{H}_1^T / H_uH_d$ & $H_1^T \bar{H}_2^T/ H_u \bar{H}_2^D$ & $H_2^T \bar{H}_1^T/H_2^D H_d $ & $H_2^T \bar{H}_2^T/ H_2^D \bar{H}_2^D$ \\
 \hline
$U(1)_{24}^A$ & 0 / 0 & $4/-6$ & $-4/6$ & 0/0 \\
$Z_{2nR}$ & 0 &$r_3-2m$ & $2-r_3$ & $2-2m$ \\
$Z_{n+1}$ &  0 & $q_3+m$ & $-q_3$ & $m$\\
\hline
\end{tabular}
\end{center}
\label{tab:chargehiggs}
\end{table}%

The $R$ symmetry is broken at the GUT scale by the VEVs of $\Phi$, $\bar{\Phi}$ and $\phi$.
After  spontaneous $R$ symmetry breaking, the mass terms of Higgs multiplets are generated via 
the couplings to those VEVs.
From the charge assignments shown in Table~\ref{tab:charge1},
the following Higgs bi-linear terms in the superpotential are allowed,
\begin{eqnarray}
\label{eq:DTW}
W = H_1 \bar{\Phi}_3 \bar{H}_2 + H_2 \Phi_3 \bar{H}_1 + \phi^m H_2 \bar{H}_2,
\end{eqnarray}
which lead to the GUT scale masses to Higgs fields except for $H_u$ and $H_d$.
It should be emphasized that the Higgs bi-linear term $H_1\bar{H}_1$ does not appear due to 
the $Z_{n+1}$ symmetry in spite of the vanishing charges of $H_1\bar{H}_1$.
On the other hand, the Higgs bi-linear term  $H_2\bar{H}_2$ proportional to $\phi^m$ is allowed
by the $Z_{n+1}$ symmetry, which leads to the mass of $O(v)$ or $O(v^2)$ for either $m=1$
or $m=2$.

The $U(1)_{24}^A$ symmetry also plays an important role in the doublet-triplet splitting.
First, let us remember that the order parameters $v_3$ and $\bar{v}_3$ have $(U(1)_{24}^A, Z_{2nR},Z_{n+1})$ 
charges of $(4,r_3,q_3)$ and $(-4,2+2m-r_3,-m-q_3)$, respectively.
Thus, the triplet mass terms $v_3 H_2^T \bar{H}_1^T$ and $\bar{v}_3 H_1^T \bar{H}_2^T$
are allowed by symmetries (see Table~\ref{tab:chargehiggs}).
The doublet mass terms such as $v_3 H_2^D H_d$ and $\bar{v}_3 H_u \bar{H}_2^D$
are on the other hand forbidden by $U(1)_{24}^A$ symmetry, although 
they can {\it never} be forbidden by $Z_{2nR}$ nor $Z_{n+1}$ symmetries.

From the above arguments, we see that both $U(1)_{24}^A$ and $Z_{n+1}$ symmetries play the roles of triplet symmetries,
where the doublet mass term $H_u H_d$ are neutral under those symmetries.
That is, due to the vanishing charges of $H_u H_d$ under the $U(1)_{24}^A$ and $Z_{n+1}$ symmetries, 
couplings to $v_3$, $\bar{v}_3$, $v_2$, $\bar{v}_2$ and $\vev{\phi}$ are highly surpressed, while 
other components in the Higgs sector obtain large masses.
The doublet mass, i.e.~the $\mu$-term is eventually generated via the Planck-suppressed interactions 
to the order parameters of the $R$ symmetry with charge $2$ but neutral under the triplet symmetries,
\begin{eqnarray}
W = \phi^{n+1} H_1\bar{H}_1\ , \quad v^n \phi H_1 \bar{H}_1 \ .
\end{eqnarray}
It should be emphasized that the scales of $\vev{\phi}^{n+1}$ and $v^n\vev\phi$ are nothing but the one of the
VEV of the superpotential, $\vev{W}$, and hence,
the doublet mass of the gravitino mass is naturally achieved as expected in the 
solution to the $\mu$-problem in the MSSM using the $R$ symmetry breaking~\cite{Inoue:1991rk,Casas:1992mk}.

Several comments are in order.
In the above discussion, we have focused on the suppression of the bi-linear term $H_1\bar H_1$.
It should be noted, however, that the mass term of $H_u H_d$ could be induced by mixing  of them with 
$H_2^D$ and $\bar{H}_2^D$. 
In fact, the $U(1)_{24}^A$ symmetry alone does not forbid mass terms proportional to
$\bar{v}_2 H_u \bar{H}_2^D $ and $v_2 H_2^D H_d $. 
In the model discussed in the next section, we have checked that mass terms of
$H_u \bar{H}_2^D $ and $H_2^D H_d $ are sufficiently suppressed due to
charge-mismatching between these bilinear terms and order parameters of the $R$ symmetry breaking.

So far, we have given GUT scale masses to Higgs the multiplets other than $H_u$ and $H_d$.
Still, it is necessary to consider masses of GUT breaking fields $\Phi_3$, $\Phi_2$, $\bar{\Phi}_3$ and $\bar{\Phi}_2$.
We discuss this topic in the next section along with the unification scale and the decay rate of protons.

\subsection{Bottom-tau unification}
\begin{figure}[t]
\begin{center}
\begin{minipage}{.45\linewidth}
  \includegraphics[width=\linewidth]{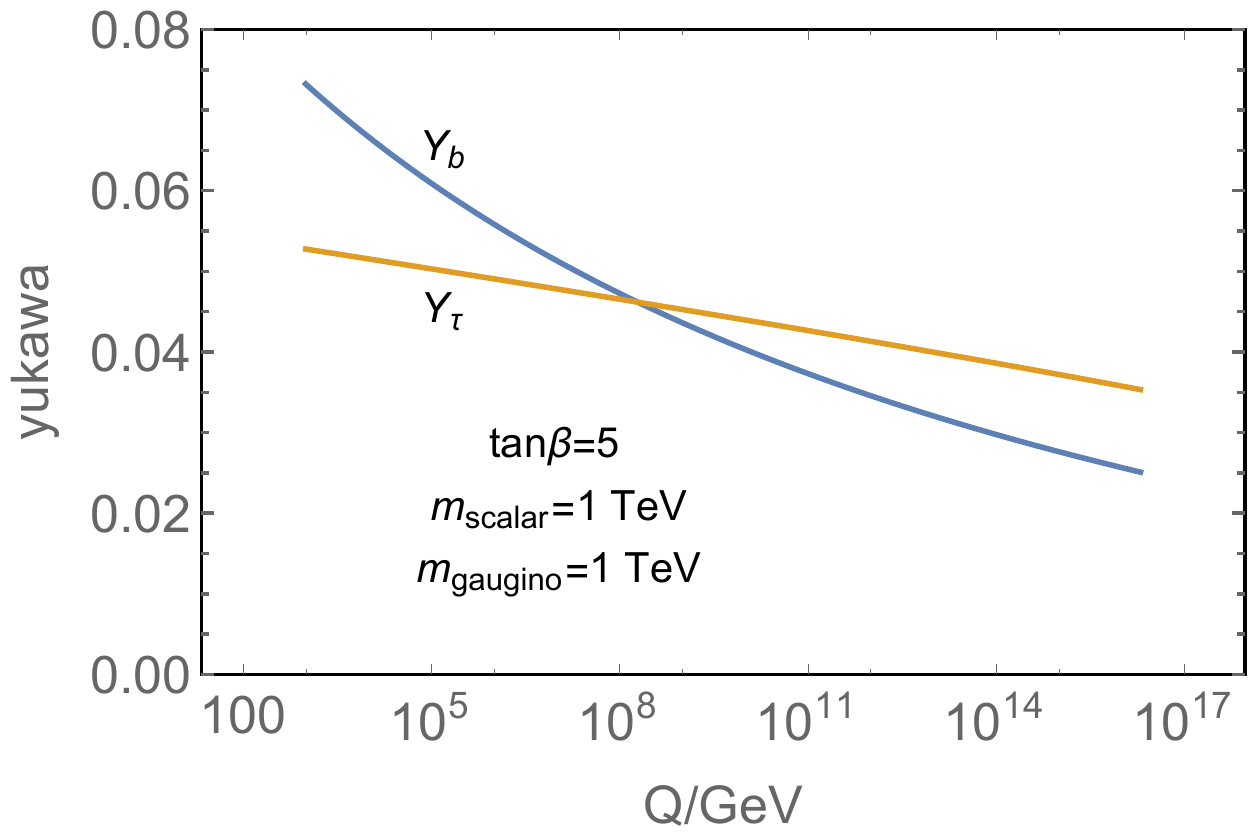}
 \end{minipage}
 \hspace{1cm}
 \begin{minipage}{.45\linewidth}
  \includegraphics[width=\linewidth]{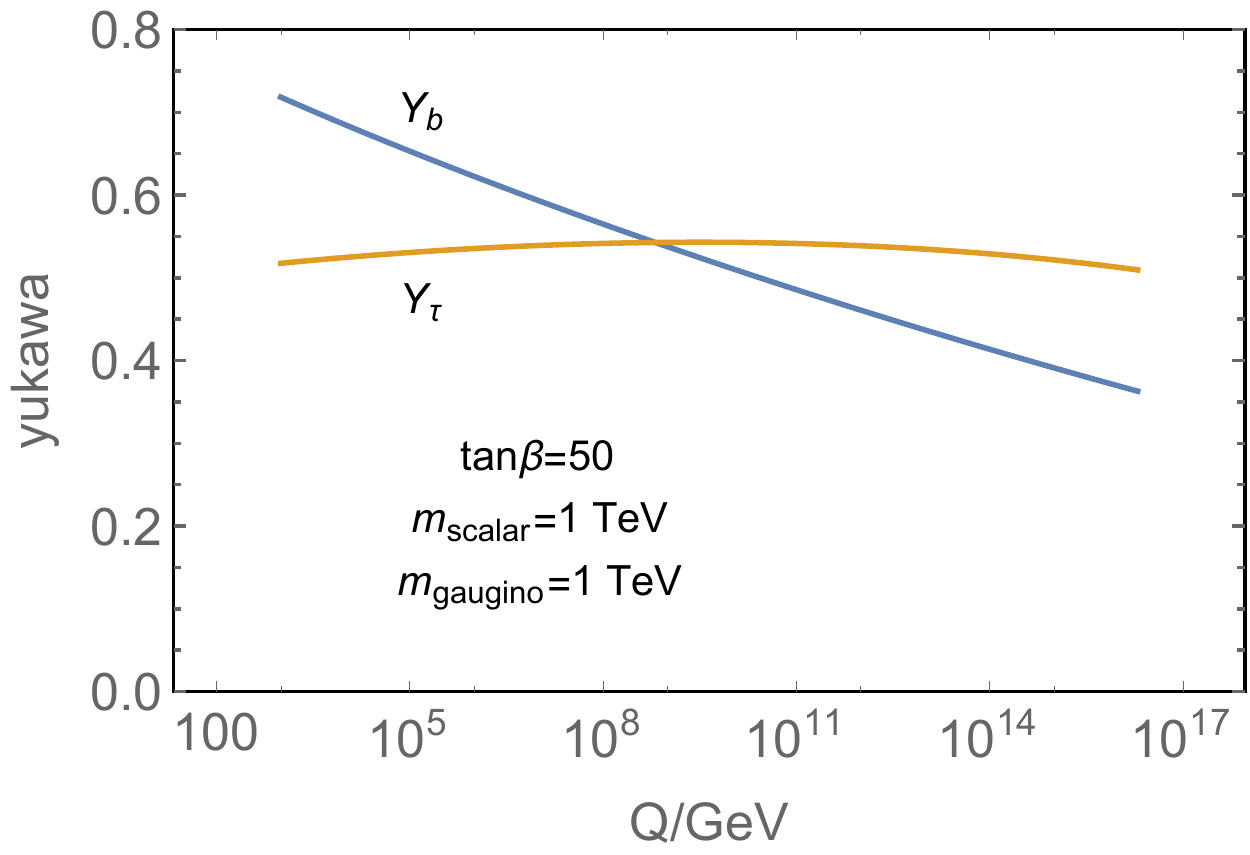}
 \end{minipage}
 \end{center}
\caption{\sl \small
Running of the bottom yukawa coupling $Y_b$ and  the tau yukawa coupling $Y_{\tau}$
for $\tan\beta = 5$ (left) and $\tan\beta = 50$(right).
The x-axis, $Q$, denotes the renormalization scale.
The discrepancy between $Y_\tau$ and $Y_b$ becomes larger for larger soft scalar masses.}
\label{fig:yukawas}
\end{figure}

Here, we briefly mention the unification of down-quark yukawa couplings and charged-lepton yukawa couplings.
As we have mentioned, we consider a model such that the MSSM doublet Higgses 
are embedded in ${\bf 5}$ and ${\bf \bar{5}}$ of $SU(5)_1$, and quarks and leptons are unified into ${\bf \bar{5}}$ and ${\bf 10}$ of $SU(5)_1$.
Then down-quark yukawa couplings and charged-lepton yukawa couplings must be unified at the GUT scale.
In Fig.~\ref{fig:yukawas}, we show the renormalization group running of the bottom yukawa coupling $Y_b$ and  the tau yukawa coupling $Y_{\tau}$ in the MSSM.
Here, we assume that SUSY particles are as heavy as $1$ TeV.
It can be seen that $Y_{\tau} > Y_{b}$  around that GUT scale, which is inconsistent with the unification of yukawa couplings.
The discrepancy becomes larger for a high scale SUSY breaking models where the gaugino masses 
are kept in the TeV region (see e.g. \cite{Giudice:2004tc}).

This problem can be easily mended by introducing ${\bf 5}$ and ${\bf \bar{5}}$ of $SU(5)_2$, which we denote as ${\bf 5}_2$ and ${\bf \bar{5}}_2$, respectively.
Consider the following superpotential,
 \begin{eqnarray}
W = M {\bf 5}_2\bar{{\bf 5}}_2+\lambda' \bar{\bf 5}\bar{\Phi}_3 {\bf 5}_2+ y \bar{H}_1 {\bf 10}~\bar{\bf 5},
 \end{eqnarray}
where $M$, $\lambda'$ and $y$ are constants.%
\footnote{
One may replace the constant $M$ with fields which obtain GUT scale VEVs.
}
This superpotential is always allowed with appropriate choice of charges of ${\bf 5}_2$ and ${\bf \bar{5}}_2$.
Here, we consider only the third generation of ${\bf 10}$ and ${\bf \bar{5}}$, for simplicity.

After the GUT breaking, the triplet in ${\bf \bar{5}}$ mixes with that in ${\bf \bar{5}}_2$.
The right-handed bottom quark we observe in low energy scales is a linear combination of them.
Just below  the GUT scale, the bottom yukawa coupling is smaller than the tau yukawa coupling
by a factor of $O(1)$, if $M = O(v_3)$.%
\footnote{A similar mechanism to split the yukawa couplings can be implemented in a simple $SU(5)$ GUT model.
}
This is consistent with the running of yukawa couplings shown in Fig.~\ref{fig:yukawas}.
Similarly, one can arrange $O(1)$ differences between down-quark yukawa couplings and charged-lepton yukawa couplings of the first and the second generations, by mixing of triplets in ${\bf \bar{5}}$ of the first and the second generations with the triplet in ${\bf \bar{5}}_2$.

\section{Mass spectrum, unification and proton decay}
\label{sec:masses}
In this section, we study models outlined in the previous section in detail.
We present a model in which masses of $\Phi_3$, $\Phi_2$, $\bar{\Phi}_3$ and $\bar{\Phi}_2$ are sufficiently large.
We show the mass spectrum and estimate unification scales as well as decay rates of the proton in the model.

\subsection{Mass spectrum}
In our model, we take $n=6$,
and hence, the discrete symmetries are  $Z_{12R}$ and $Z_7$.
In Table~\ref{tab:matter1}, we show the matter content and the charge assignment.
With this charge assignment, the masses of Higgs multiplets are given by the one in previous section with $m=2$.
In addition to the fields introduced in the previous section, we introduce
gauge singlet fields $\bar\phi$, $Z_1$, $Z_2$, $Z_3$, $Z_4$ and a spurious field $\bar{v}^6$.
We assume that $v^6 \sim \bar{v}^6$, which is natural since they have opposite charges with each other.

\begin{table}[t]
\caption{\small \sl
Matter contents and charge assignment.} 
\label{tab:matter1}
\small{
\begin{center}
\begin{tabular}{|c|cccccccccccc|}
\hline
 & $\Phi_{3}$& $\bar{\Phi}_{3}$ 
  & $\Phi_{2}$& $\bar{\Phi}_{2}$
  & $\phi$& $\bar\phi$
  & $v^6$& $\bar{v}^6$ &   $Z_1$ 
  & $Z_2$& $Z_3$&$Z_4$
  \\
 \hline
$Z_{12R}$ & $-4$ &$-2$ &$6$ &$0$ 
& $2$& $2$& $0$& $0$ &    $-2$ & $-2$&$6$&$2$
\\
$Z_{7}$ & $0$ &$2$ &$4$&$-2$ 
& $1$& $-1$& $-1$& $1$ &   $1$ & $1$&$2$&$3$\\
\hline
\end{tabular}

\begin{tabular}{|c|cccccc|}
\hline
  & $H_1$& $\bar{H}_1$
  & $H_2$& $\bar{H}_2$
  & ${\bf \bar{5}}$& ${\bf 10}$\\
 \hline
$Z_{12R}$ & $0$& $0$& $6$& $-8$& $1$& $1$
\\
$Z_{7}$ &  $0$& $0$& $0$& $-2$ & $0$& $0$\\
\hline 
\end{tabular}
\end{center}}
\end{table}

The superpotential
which ensure the VEV pattern in Eq.~(\ref{eq:VEV}) is given by,
\begin{eqnarray}
W &=& \bar\phi(\Phi_3\Phi_3\Phi_3\Phi_2\Phi_2+\bar{v}^6) 
  \nonumber \\
 &&+Z_3(\Phi_3\bar{\Phi}_2)+(\Phi_3\bar{\Phi}_2\Phi_3\bar{\Phi}_3)+(\Phi_3\bar{\Phi}_2\Phi_2\bar{\Phi}_2)+(\Phi_3\bar{\Phi}_2)(\Phi_3\bar{\Phi}_3)+(\Phi_3\bar{\Phi}_2)(\Phi_2\bar{\Phi}_2), \nonumber \\
& &+(\Phi_2\bar{\Phi}_3\Phi_2\bar{\Phi}_3\Phi_3\bar{\Phi}_3)+(\Phi_2\bar{\Phi}_3\Phi_2\bar{\Phi}_3\Phi_2\bar{\Phi}_2)
+(\Phi_3\Phi_3\Phi_2\Phi_2\Phi_2)\phi^2
\nonumber\\
&& + Z_1(\Phi_2\bar{\Phi}_3)+(Z_1+Z_2)(\bar{\Phi}_3\bar{\Phi}_3\bar{\Phi}_3\bar{\Phi}_3\bar{\Phi}_2 )
\nonumber\\
&&+ Z_4Z_3(\Phi_3\bar{\Phi}_3) + Z_4Z_3(\Phi_2\bar{\Phi}_2) + Z_4(\Phi_3\bar{\Phi}_3)(\Phi_3\bar{\Phi}_3) + Z_4(\Phi_2\bar{\Phi}_2)(\Phi_3\bar{\Phi}_3) 
\nonumber\\
&&+ Z_4(\Phi_2\bar{\Phi}_2)(\Phi_2\bar{\Phi}_2)+ Z_4(\Phi_3\bar{\Phi}_3\Phi_3\bar{\Phi}_3)+Z_4(\Phi_2\bar{\Phi}_2\Phi_2\bar{\Phi}_2)
\label{eq:GUT breaking1}
\end{eqnarray}
where we  omit coupling constants, which we assume to be $O(1)$ in the Planck unit.
Here,
\begin{eqnarray}
(\Phi_3\Phi_3\Phi_3\Phi_2\Phi_2) \equiv \epsilon^{abcde}\epsilon_{ABCDE}\Phi_{3a}^{\ A}\Phi_{3b}^{\ B}\Phi_{3c}^{\ C}\Phi_{2d}^{\ D}\Phi_{2e}^{\ E},~~
(\Phi \bar{\Phi}\cdots \bar{\Phi}) \equiv \Phi_a^A \bar{\Phi}_A^b \cdots \bar{\Phi}_C^a,
\end{eqnarray}
where
lower indices $a,b,\cdots (=1\mathchar`-5)$ and $A,B,\cdots (=1\mathchar`-5)$ are indices of the fundamental representation of $SU(5)_1$ and $SU(5)_2$, respectively. Upper indices are those of the anti-fundamental representation.
Here, we have shown only relevant terms for the later discussion and omitted several terms which are allowed by symmetries.
Our conclusions are not changed, even if we includes those terms.

The $F$ term condition of $\bar\phi$ requires $\Phi_3$ and $\Phi_2$ to obtain their VEVs.%
\footnote{Since $\bar\phi^7$ is allowed by symmetry, there is a gauge symmetric where $\bar\phi\sim \bar{v}$ and $\Phi=\bar{\Phi}=0$,
in addition to the  $SU(5)_1\times SU(5)_2$ branch, i.e. $\Phi=\bar{\Phi}\neq 0$.
We choose the symmetry breaking branch.
}
Along with the $F$ term condition of $\bar\phi$ and the $D$ term condition, the $F$ term conditions of $Z_{1,2,4}$ and
$\Phi$'s 
require the VEVs of the form in Eq.~(\ref{eq:VEV}) with $v_3=\bar{v}_3 \sim v_2 = \bar{v}_2 \equiv v_G \sim \bar{v}^{6/5}$, 
$\vev{Z_1} \sim v^2 v_G^3$, $\vev{Z_2} \sim v^2$ and $\vev{Z_3} \sim v^2$.%
\footnote{It should be noted that  superpotential terms such as $\phi^4 Z_{1,2}^3$ are also allowed by symmetries.
In the presence of those terms, the missing VEV pattern in Eq.\,(\ref{eq:VEV}) is destabilized, although $SU(3)_c\times SU(2)_L\times U(1)_Y$ remains unbroken.
Such destabilization potentially leads to too large doublet Higgs mass and/or too large VEV of the superpotetntial.
In our model,  we have checked that the missing  part of the VEV of $\Phi$'s are destabilized by 
$\delta \Phi_3 = O(v^{10}/v_G^3)$ and $\delta \bar{\Phi}_3  =O(v^8/v_G^4)$, 
which leads to the doublet mass of $O(v^{16}/v_G^7)$ and the VEV of the superpotential of $O(v^{10})$.
Thus, those terms causes no serious problems.
}
Here, the singlet fields $Z_{1,2}$ cancel the $F$ terms of $\Phi$'s.
Note that the VEVs of $\Phi$ and $\bar{\Phi}$ are as large as $\vev{\phi}$.
Thus, the origin of the GUT scale and the gravitino mass are interrelated 
with each other, which is an interesting feature of this example.

Let us consider the masses of $SU(3)_c\times SU(2)_L \times U(1)_Y$ charged particles in $\Phi_3$, $\Phi_2$, $\bar{\Phi}_3$ and $\bar{\Phi}_2$.
In the unitarity gauge, their charged components are decomposed as
\begin{eqnarray}
\Phi_3 =
\begin{pmatrix}
v_3 + O_3 & X \\
Y & T_3
\end{pmatrix},
\bar{\Phi}_3 =
\begin{pmatrix}
\bar{v}_3 + O_3 & \bar{X} \\
\bar{Y} & \bar{T}_3
\end{pmatrix},
\Phi_2 =
\begin{pmatrix}
O_2 & \bar{X} \\
\bar{Y} & v_2 + T_2
\end{pmatrix},
\bar{\Phi}_2 =
\begin{pmatrix}
\bar{O}_2 & X \\
Y & \bar{v}_2 + T_2
\end{pmatrix},
\end{eqnarray}
where $SU(3)_c\times SU(2)_L \times U(1)_Y$ charges are given by
\begin{eqnarray}
O : ({\bf 8} , {\bf 1})_0,~~T: ({\bf 1}, {\bf 3})_0,~~X: ({\bf 3},{\bf 2})_{-5/6},~~Y: ({\bf \bar{3}},{\bf 2})_{5/6} .
\end{eqnarray}
The second lines of superpotential in Eq.~(\ref{eq:GUT breaking1}) give the masses of $O(v_G^2)$ to
$O_3\bar{O}_2$, $T_3 T_2$ and $XY$. 
The masses of, $O_2$, $\bar{X}$, $\bar{Y}$ and $\bar{T}_3$ are generated from the third line of the 
superpotential.
The first and the second term in the third line give $O(v_G^4)$ masses to $O_2$ and $\bar{T}_3$, respectively.
The third term of the third line gives an $O(v_G^3 v^2) \sim O(v_G^{14/3})$ mass to $\bar{X}~\bar{Y}$.

Let us summarize masses of charged particles lighter than the GUT scale $v_G$:
\begin{eqnarray}
\label{eq:massM1}
O(v_G^{5/3})&:&
H_2^D,~\bar{H}_2^D \nonumber \\
O(v_G^2)&:&
~O_3,~\bar{O}_2,~T_3,~T_2,~X,~Y \nonumber \\
O(v_G^4)&:&
 O_2,~\bar{T}_3 \nonumber \\
 O(v_G^{14/3})&:& \bar{X},~\bar{Y}.
\end{eqnarray}
Note that relatively light fields, $O_2$, $\bar{X}$, $\bar{Y}$ and $\bar{T}_3$, form a GUT complete multiplet ${\bf 24}$
in terms of a single $SU(5)$ group, and hence, the GUT unification is maintained.
We have also checked that all the gauge singlet fields below the GUT scale obtain sizable masses.
Since all charged fields obtain masses, the symmetry breaking branch is found to be a stable vacuum.
Besides, since all the masses are larger than the size of the (gravity mediated) SUSY breaking soft masses
of $O(v^7)$, the vacuum is even stable against the SUSY breaking effects.

\subsection{Unification scale and proton decay}
\subsubsection*{Unification scale}
\begin{figure}[t]
\begin{center}
\begin{minipage}{.45\linewidth}
  \includegraphics[width=\linewidth]{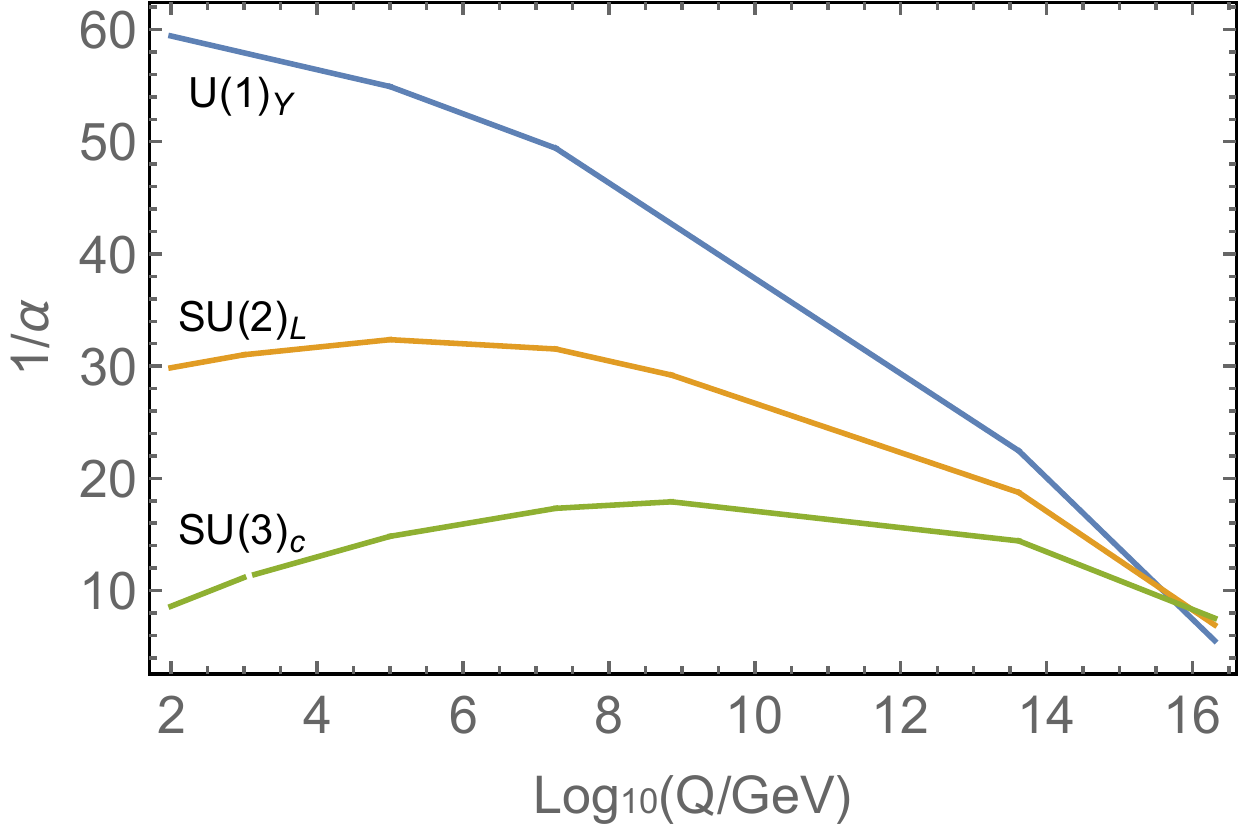}
 \end{minipage}
 \hspace{1cm}
 \begin{minipage}{.45\linewidth}
  \includegraphics[width=\linewidth]{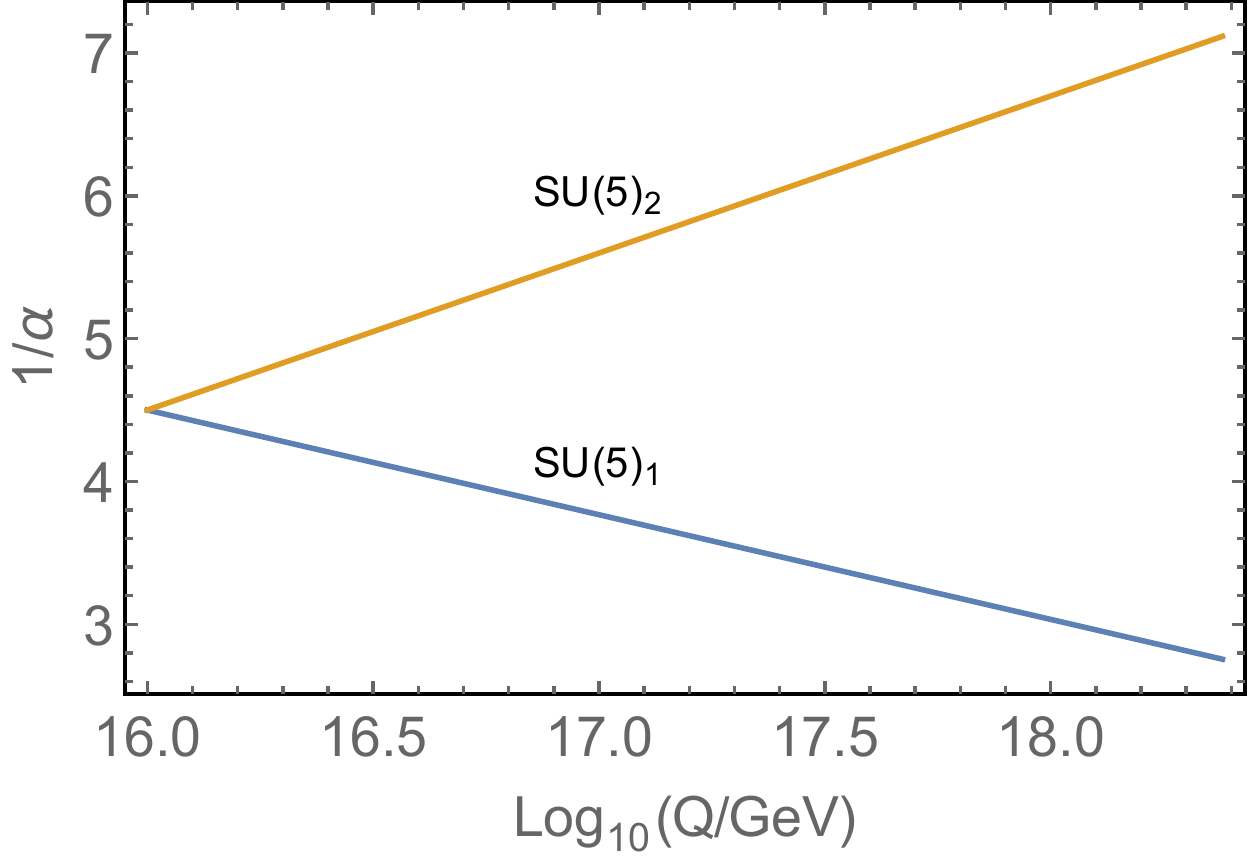}
 \end{minipage}
  \end{center}
\caption{\sl \small
Running of the gauge coupling constants.
We assume that $v_3= v_2 = 2\times 10^{16}$\,GeV, MSSM gaugino masses are $1$\,TeV, MSSM scalar masses are 
$100$\,TeV, and the Higgsino mass is $100$\,TeV.
The left panel show the running in the low energy effective theory below the GUT scale.
The right panel show the running above the GUT scale.
}
\label{fig:gauge run 1}
\end{figure}

In the left panel of Fig.~\ref{fig:gauge run 1}, we show the one-loop renormalization group running of the gauge coupling constants of $SU(3)_c$, $SU(2)_L$ and $U(1)_Y$.
Here, we assume that $v_3= v_2 = 1\times 10^{16}$ GeV
to determine the masses of charged particles.
We also assume that the MSSM gaugino masses are $1$ TeV, the MSSM scalar masses are $100$ TeV, and the Higgsino mass is $100$ TeV, which is the case of typical high scale SUSY breaking models.
The masses of the heavy fields in Eq.\,(\ref{eq:massM1}) are summarized in Table\,\ref{tab:mass}.

\begin{table}
\caption{\sl \small Mass spectrum of charged particles.
We assume  $v^6 = \bar{v}^{6}$, $v_G \sim v^{5/6}$.
In the parentheses, we take the gravitino masses $100$\,TeV.} 
\label{tab:mass}
\begin{center}
\begin{tabular}{|c|c||c|c||c|c||c|ccccccc}
\hline
  &  {\rm Mass}\,(GeV)& &  {\rm Mass}\,(GeV)&& {\rm Mass}\,(GeV)
  \\
 \hline
 \hline
$({\bf 8},{\bf 1})_0$ & $v_G^2\ (10^{14})$  &$({\bf 1},{\bf 3})_0$ & $v_G^2\ (10^{14})$& $({\bf 3},{\bf 2})_{-6/5},({\bf \bar{3}},{\bf 2})_{6/5}$ & $v_G^2\ (10^{14})$\\
 \hline
$({\bf 8},{\bf 1})_0$ & $v_G^2\ (10^{14})$&$({\bf 1},{\bf 3})_0$ & $v_G^2\ (10^{14})$&$({\bf 3},{\bf 2})_{-6/5},({\bf \bar{3}},{\bf 2})_{6/5}$ & $v_G^{14/3}\ (10^{8})$\\
 \hline
$({\bf 8},{\bf 1})_0$ & $v_G^4\ (10^{10})$&$({\bf 1},{\bf 3})_0$ & $v_G^4\ (10^{10})$& $({\bf 1},{\bf 2})_{1/2},({\bf \bar{1}},{\bf 2})_{-1/2}$   & $v_G^{5/3}\ (10^{14})$ \\
\hline

\end{tabular}
\end{center}
\end{table}

In the right panel of Fig.~\ref{fig:gauge run 1},
we show the one-loop renormalization group running of the gauge couplings constants of $SU(5)_1$ and $SU(5)_2$ from $1\times 10^{16}$ GeV to the Planck scale.
We take $g_{SU(5)_1} = g_{SU(5)_2}$ at $1\times 10^{16}$ GeV.
Here, we include the contribution of ${\bf 5}_2$ and ${\bf \bar{5}}_2$ which are introduced to solve the problem of the unification of yukawa couplings.
One can see that both gauge coupling constants are perturbative below the Planck scale.%
\footnote{Since $SU(5)_2$ is asymptotically free, we might safely assume $g_{SU(5)_1}\ll g_{SU(5)_2}$
without spoiling the perturbativity up to the Planck scale.}

\subsubsection*{Proton decay}
\begin{figure}[t]
\begin{center}
  \includegraphics[width=0.5\linewidth]{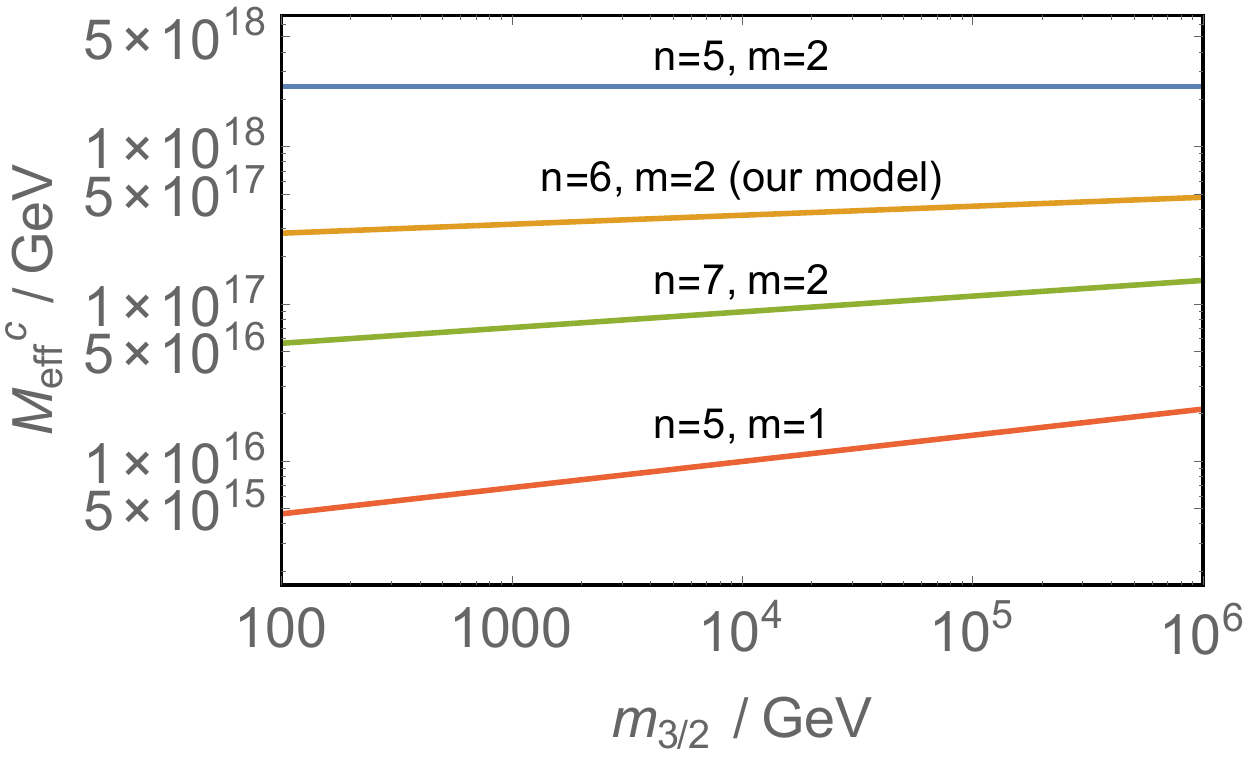}
   \end{center}
\caption{\sl \small
The effective triplet mass $M_{\rm eff}^c$ for a given gravitino mass $m_{3/2}$.
}
\label{fig:meffc}
\end{figure}

Now, we are at the position to predict  the proton lifetime.
Let us first consider the proton decay by dimension five operators~\cite{dimension5}.
The Planck-suppressed dimension five operators such as,
\begin{eqnarray}
W = {\bf 10} \,{\bf 10}\, {\bf 10}\, {\bf \bar{5}},
\end{eqnarray}
are forbidden due to the discrete $R$ symmetry, i.e. the doublet symmetry.
However, below the GUT scale, the exchange of the triplets $H_1^c$ and $\bar{H}_1^c$ induces a dimension five operator:
\begin{eqnarray}
\label{eq:effdim5}
W &=&H_1\bar{\Phi}_3\bar{H}_2+\bar{H}_1\Phi_3H_2+\phi^mH_2\bar{H}_2+ y_{IJ} H_1 {\bf 10}^I{\bf 10}^J +y'_{iI} \bar{H}_1 {\bf \bar{5}}^i {\bf 10}^I \nonumber \\
&\rightarrow& \frac{1}{M_{\rm eff}^c} y_{IJ} y'_{iK} Q^IQ^JQ^KL^i + \cdots,~~M_{\rm eff}^c \equiv \frac{v_3 \bar{v}_3}{\vev{\phi}^m},
\end{eqnarray}
where $i,I (= 1\hyphen 3)$ denote generation indices.
The decay rate of $p\rightarrow K^+ \bar{\nu}$ induced by the dimension five operator is roughly given by
\begin{eqnarray}
\Gamma(p \rightarrow K^+ \bar{\nu})^{-1} \approx 10^{39}~{\rm years} \times {\rm sin}^2 2\beta \left(\frac{M_{\rm SUSY}}{100\,{\rm TeV}}\right)^2 \left(\frac{M_{\rm eff}^c}{10^{18}\,{\rm GeV}}\right)^2,
\end{eqnarray}
where $M_{\rm SUSY}$ denotes masses of SUSY particles.
In Fig.~\ref{fig:meffc}, we show the relation between the gravitino mass $m_{3/2}$ and the effective triplet mass $M_{\rm eff}^c$.%
\footnote{Although $M_{\rm eff}^c$ exceeds the Planck scale for some parameters, the contributions 
from Eq.\,(\ref{eq:effdim5}) still dominates since the Planck suppressed operators of the dimension five proton decay
are forbidden by symmetries.} 
The figure shows that $M_{\rm eff}^c$ is sufficiently large and hence the models are 
consistent with the current 90$\%$ C.L.~limit,
$\Gamma(p \rightarrow K^+ \bar{\nu}_\mu)^{-1} > 4.0\times 10^{33}$ years~\cite{Abe:2011ts}.

It should be noted that the $\mu$-term and the dimension five proton decay operators are closely
related in terms of symmetry.
Assuming the standard embedding of the Higgs multiplets and the quarks and the leptons,
the charge assignment of the $\mu$-term ($Q_{\mu}$) under a low energy symmetry is 
related to that of the dimension five operators ($Q_{\rm dim 5}$) by $Q_{\mu}= Q_{\rm dim 5}$
for non-$R$ symmetries and $Q_{\mu} = 4 - Q_{\rm dim 5}$ for $R$ symmetry.
Thus, the absence of the $\mu$-term by a symmetry means the absence of  
dimension five proton decay operators.
In our model, however, the doublet symmetry which leads to an effective low energy symmetry
forbidding the $\mu$-term is spontaneously broken at the GUT scale.
Therefore, the dimension five proton decay operators can be generated even if the doublet Higgs mass 
is suppressed by the triplet symmetries.

Next, we consider the proton decay by dimension six operators, namely by the exchange of GUT gauge bosons~\cite{Georgi:1974sy,Georgi:1974yf}.
The decay rate of $p \rightarrow \pi^0+e^+$ by dimension six operators is roughly given by,
\begin{eqnarray}
\Gamma(p \rightarrow \pi^0+e^+)^{-1} \approx 10^{35}~{\rm years} \left(\frac{v_G}{ 10^{16} {\rm GeV}}\right)^4,
\end{eqnarray}
where $v_G$
is the vacuum expectation value of GUT breaking fields, $v_G \sim v_3 \sim v_2$.
In our model, the decay rate is consistent with the current 90$\%$ C.L.~limit,
$\Gamma(p \rightarrow \pi^0+e^+)^{-1} > 1.3 \times10^{34}$ years~\cite{Abe:2011ts}.

\section{Conclusions and Discussion}
In this paper, we have revisited the long-standing problem of SUSY GUT models, the 
doublet-triplet splitting problem. 
We payed particular attention to the consistency of
the low energy symmetry (doublet symmetry) which controls the size of the $\mu$-term
to GUT models.
By using the anomaly matching conditions of the doublet symmetry, we find that 
the low energy symmetry cannot be embedded in GUT models with automatic 
gauge coupling unification unless the low energy symmetry is broken at the GUT scale.
It should be noted that the above no-go theorem applies even to GUT models based 
on product gauge groups as long as the coupling unification is automatic.

We also proposed a new prescription to embed the low energy doublet symmetry to the GUT models
while evading the no-go theorem.
There, the doublet symmetry is spontaneously broken at the GUT scale.
Since order parameter of the doublet symmetry are also charged under other symmetries (we call them triplet symmetries),
the doublet Higgses do not obtain a mass of the GUT scale.
The triplet Higgses (and other unwanted fields), on the other hand, are charged under the
triplet symmetry so that they obtain much heavier masses than the doublet Higgses.
As a notable feature of this mechanism, the less charged fields obtain more suppress masses, i.e., the haves 
get large masses while the have-nots get no masses at the GUT scale.

We also found a similar no-go theorem on the $R$ symmetry, so that $R$ symmetry should be broken
at the GUT scale.
There again, the $R$ symmetry 
is effectively preserved as a good approximate symmetry below the GUT scale,
if the order parameters are charged under other symmetries.

As a demonstration, we consider an $SU(5)_1\times SU(5)_2$ GUT model 
where the doublet symmetry is $R$ symmetry. 
The $\mu$-term is generated via the coupling of the doublet Higgses
to $R$ symmetry breaking of the order of the VEV of the superpotential,
and hence, the doublet Higgs obtains the mass of the order of the gravitino mass. 
This shows that the successful MSSM such as the pure gravity mediation model
 can be consistently extended
to GUT models with an automatic coupling unification.

One of the interesting observations of this example is a connection between
the GUT scale and the size of the gravitino mass.
As we discussed in Sec.~\ref{sec:masses}, the origins of the GUT scale and the scale of the spontaneous 
$R$ symmetry breaking is naturally unified while the VEV of the superpotential (i.e.~the gravitino mass)
is much suppressed due to triplet symmetries.
Thus, for a given GUT scale, the required SUSY breaking scale to obtain a flat-universe by fine-tuning is 
no more free parameters.
This feature might shed light on the question why the SUSY breaking scale is not in a range 
such that
the electroweak symmetry breaking is naturally induced by the SUSY breaking, although
we have no definite answers on this question.

Several comments are in order.
In many models, the low-energy symmetry which controls the size of the $\mu$-term
is not anomaly-free.%
\footnote{Here, we now discuss the total anomaly including not only the Higgs sector
but also gauginos, quarks and leptons.}
Thus, those symmetries are thought to be difficult to originate from exact gauge symmetries.
In our mechanism, on the other hand, the doublet symmetry is spontaneously broken
at the GUT scale.
Hence,
the failure in the anomaly free conditions at the low energy
does not necessarily mean that the low-energy symmetry cannot originate
from exact gauge symmetries. 

The low-energy symmetry which suppresses the $\mu$-term also suppresses
the dimension five proton decay operators.
In our mechanism, however, the doublet symmetry is broken at the GUT scale, and hence,
the dimension five proton decay operators are not necessarily highly suppressed.
In fact, as we have discussed in Sec.~\ref{sec:masses}, the sizable (but acceptable) dimension five proton decay 
operators appear.
It should be noted that the relative sizes of the dimension six proton decay operators and the 
dimension five proton decay operators can be far different from the minimal (and fine-tuned) $SU(5)$
GUT model, with which we might be able to distinguish our mechanisms from the minimal $SU(5)$ model.

%
%

\section*{Acknowledgements}
This work is supported by Grant-in-Aid for Scientific research from
the Ministry of Education, Culture, Sports, Science
and Technology (MEXT) in Japan,
No. 24740151 
and 25105011 (M.I.), from the Japan Society for the Promotion of Science (JSPS), No. 26287039 (M.I.),
the World Premier International Research Center Initiative (WPI Initiative), MEXT, Japan (M.I. and M.S.),
and a JSPS Research Fellowships for Young Scientists (K.H.).


\begin{thebibliography}{99}

\bibitem{Maiani:1979}
L. Maiani, in Summer School on Particle Physics, Paris, France (1979).

\bibitem{Veltman:1980mj} 
  M.~J.~G.~Veltman,
  Acta Phys.\ Polon.\ B {\bf 12}, 437 (1981).

\bibitem{Witten:1981kv} 
  E.~Witten,
  Phys.\ Lett.\ B {\bf 105}, 267 (1981).

\bibitem{Kaul:1981wp} 
  R.~K.~Kaul,
  Phys.\ Lett.\ B {\bf 109}, 19 (1982).


\bibitem{Georgi:1974sy} 
  H.~Georgi and S.~L.~Glashow,
  Phys.\ Rev.\ Lett.\  {\bf 32}, 438 (1974).
  \bibitem{lhc_susy}
  S.~Chatrchyan {\it et al.}  [CMS Collaboration],
  JHEP {\bf 1406}, 055 (2014)
  [arXiv:1402.4770 [hep-ex]];
  G.~Aad {\it et al.}  [ATLAS Collaboration],
  JHEP {\bf 1409}, 176 (2014)
  [arXiv:1405.7875 [hep-ex]].
    
  \bibitem{lhc_higgs}
  G.~Aad {\it et al.}  [ATLAS Collaboration],
  Phys.\ Lett.\ B {\bf 716}, 1 (2012)
  [arXiv:1207.7214 [hep-ex]];
  Phys.\ Rev.\ D {\bf 90}, no. 5, 052004 (2014)
  [arXiv:1406.3827 [hep-ex]].
  S.~Chatrchyan {\it et al.}  [CMS Collaboration],
  Phys.\ Lett.\ B {\bf 716}, 30 (2012)
  [arXiv:1207.7235 [hep-ex]];
  Phys.\ Rev.\ D {\bf 89}, no. 9, 092007 (2014)
  [arXiv:1312.5353 [hep-ex]];
  V.~Khachatryan {\it et al.}  [CMS Collaboration],
  Eur.\ Phys.\ J.\ C {\bf 74}, no. 10, 3076 (2014)
  [arXiv:1407.0558 [hep-ex]].






\bibitem{Dimopoulos:1981zb} 
  S.~Dimopoulos and H.~Georgi,
  Nucl.\ Phys.\ B {\bf 193}, 150 (1981).



\bibitem{Sakai:1981gr} 
  N.~Sakai,
  Z.\ Phys.\ C {\bf 11}, 153 (1981).

\bibitem{Masiero:1982fe} 
  A.~Masiero, D.~V.~Nanopoulos, K.~Tamvakis and T.~Yanagida,
  Phys.\ Lett.\ B {\bf 115}, 380 (1982).

\bibitem{Grinstein:1982um} 
  B.~Grinstein,
  Nucl.\ Phys.\ B {\bf 206}, 387 (1982).

\bibitem{missvev}
 S.~Dimopoulos and F.~Wilczek, NSF-ITP-82-07

\bibitem{Yanagida:1994vq} 
  T.~Yanagida,
  Phys.\ Lett.\ B {\bf 344}, 211 (1995)
  [hep-ph/9409329].

\bibitem{Kawamura:2000ev} 
  Y.~Kawamura,
  Prog.\ Theor.\ Phys.\  {\bf 105}, 999 (2001)
  [hep-ph/0012125].
\bibitem{Hall:2001pg} 
  L.~J.~Hall and Y.~Nomura,
  Phys.\ Rev.\ D {\bf 64}, 055003 (2001)
  [hep-ph/0103125].

\bibitem{Izawa:1997he} 
  K.~I.~Izawa and T.~Yanagida,
  Prog.\ Theor.\ Phys.\  {\bf 97}, 913 (1997)
  [hep-ph/9703350].
  
  \bibitem{Inoue:1991rk} 
  K.~Inoue, M.~Kawasaki, M.~Yamaguchi and T.~Yanagida,
  Phys.\ Rev.\ D {\bf 45}, 328 (1992).


\bibitem{Casas:1992mk} 
  J.~A.~Casas and C.~Munoz,
  Phys.\ Lett.\ B {\bf 306}, 288 (1993)
  [hep-ph/9302227].
\bibitem{Kim:1983dt} 
  J.~E.~Kim and H.~P.~Nilles,
  Phys.\ Lett.\ B {\bf 138}, 150 (1984).


\bibitem{Goodman:1985bw} 
  M.~W.~Goodman and E.~Witten,
  Nucl.\ Phys.\ B {\bf 271}, 21 (1986).

\bibitem{Witten:2001bf} 
  E.~Witten,
  hep-ph/0201018.

\bibitem{Fallbacher:2011xg} 
  M.~Fallbacher, M.~Ratz and P.~K.~S.~Vaudrevange,
  Phys.\ Lett.\ B {\bf 705}, 503 (2011)
  [arXiv:1109.4797 [hep-ph]].
  
  

\bibitem{Izawa:1997br} 
  K.~I.~Izawa and T.~Yanagida,
  Prog.\ Theor.\ Phys.\  {\bf 99}, 423 (1998)
  [hep-ph/9710218].

\bibitem{Kitano:2001ie} 
  R.~Kitano and N.~Okada,
  Phys.\ Rev.\ D {\bf 64}, 055010 (2001)
  [hep-ph/0105220].

\bibitem{Antusch:2014poa} 
  S.~Antusch, I.~de Medeiros Varzielas, V.~Maurer, C.~Sluka and M.~Spinrath,
  JHEP {\bf 1409}, 141 (2014)
  [arXiv:1405.6962 [hep-ph]].
  



\bibitem{Dine:2002se} 
  M.~Dine, Y.~Nir and Y.~Shadmi,
  Phys.\ Rev.\ D {\bf 66}, 115001 (2002)
  [hep-ph/0206268].


  
\bibitem{Giudice:1988yz} 
  G.~F.~Giudice and A.~Masiero,
  Phys.\ Lett.\ B {\bf 206}, 480 (1988).
  
\bibitem{Ibe:2006de} 
  M.~Ibe, T.~Moroi and T.~T.~Yanagida,
  Phys.\ Lett.\ B {\bf 644}, 355 (2007)
  [hep-ph/0610277].
\bibitem{Ibe:2011aa} 
  M.~Ibe and T.~T.~Yanagida,
  Phys.\ Lett.\ B {\bf 709}, 374 (2012)
  [arXiv:1112.2462 [hep-ph]].
\bibitem{Ibe:2012hu} 
  M.~Ibe, S.~Matsumoto and T.~T.~Yanagida,
  Phys.\ Rev.\ D {\bf 85}, 095011 (2012)
  [arXiv:1202.2253 [hep-ph]].
\bibitem{Bhattacherjee:2012ed} 
  B.~Bhattacherjee, B.~Feldstein, M.~Ibe, S.~Matsumoto and T.~T.~Yanagida,
  Phys.\ Rev.\ D {\bf 87}, no. 1, 015028 (2013)
  [arXiv:1207.5453 [hep-ph]].
  
\bibitem{AMSB}
  G.~F.~Giudice, M.~A.~Luty, H.~Murayama and R.~Rattazzi,
  JHEP {\bf 9812}, 027 (1998)
  [hep-ph/9810442];
  L.~Randall and R.~Sundrum,
  Nucl.\ Phys.\ B {\bf 557}, 79 (1999)
  [hep-th/9810155];
  M.~Dine and D.~MacIntire,
  Phys.\ Rev.\ D {\bf 46}, 2594 (1992)
  [hep-ph/9205227];


\bibitem{Bagger:1999rd} 
  J.~A.~Bagger, T.~Moroi and E.~Poppitz,
  JHEP {\bf 0004}, 009 (2000)
  [hep-th/9911029].

\bibitem{D'Eramo:2013mya} 
  F.~D'Eramo, J.~Thaler and Z.~Thomas,
  JHEP {\bf 1309}, 125 (2013)
  [arXiv:1307.3251].
\bibitem{Harigaya:2014sfa} 
  K.~Harigaya and M.~Ibe,
  Phys.\ Rev.\ D {\bf 90}, no. 8, 085028 (2014)
  [arXiv:1409.5029 [hep-th]].
  
\bibitem{Coughlan:1983ci} 
  G.~D.~Coughlan, W.~Fischler, E.~W.~Kolb, S.~Raby and G.~G.~Ross,
  Phys.\ Lett.\ B {\bf 131}, 59 (1983).
\bibitem{Ibe:2006am} 
  M.~Ibe, Y.~Shinbara and T.~T.~Yanagida,
  Phys.\ Lett.\ B {\bf 639}, 534 (2006)
  [hep-ph/0605252].
\bibitem{Harigaya:2013ns} 
  K.~Harigaya, M.~Ibe, K.~Schmitz and T.~T.~Yanagida,
  Phys.\ Lett.\ B {\bf 721}, 86 (2013)
  [arXiv:1301.3685 [hep-ph]].
\bibitem{Barbieri:1994jq} 
  R.~Barbieri, G.~R.~Dvali and A.~Strumia,
  Phys.\ Lett.\ B {\bf 333}, 79 (1994)
  [hep-ph/9404278].
\bibitem{dimension5} 
 N.~Sakai and T.~Yanagida,
Nucl. Phys. B197, 533 (1982).
  S.~Weinberg,
  Phys.\ Rev.\ D {\bf 26}, 287 (1982).
  
\bibitem{Giudice:2004tc} 
  G.~F.~Giudice and A.~Romanino,
  Nucl.\ Phys.\ B {\bf 699}, 65 (2004)
  [Nucl.\ Phys.\ B {\bf 706}, 65 (2005)]
  [hep-ph/0406088].
  
\bibitem{proton decay} 
    J. ~R. ~Ellis, M. ~K. ~Gaillard, D. ~V.~ Nanopoulos and S. ~Rudaz,
    Nucl. \ Phys.\  B{\bf 176}, 61 (1980).

\bibitem{Georgi:1974yf} 
  H.~Georgi, H.~R.~Quinn and S.~Weinberg,
  Phys.\ Rev.\ Lett.\  {\bf 33}, 451 (1974).

\bibitem{Abe:2011ts} 
  K.~Abe, T.~Abe, H.~Aihara, Y.~Fukuda, Y.~Hayato, K.~Huang, A.~K.~Ichikawa and M.~Ikeda {\it et al.},
  arXiv:1109.3262 [hep-ex].
  

  



\end{thebibliography}
\end{document}